\documentclass[letterpaper,10pt]{article}
\usepackage{graphicx}
\usepackage{cite}
\usepackage{bbm}
\usepackage{amssymb}

\title{Towards understanding ELM mitigation by resonant magnetic perturbations in  MAST}

\author{ IT Chapman, A Kirk, CJ Ham, JR Harrison, YQ Liu,\\ S Saarelma, R Scannell, AJ Thornton, and the MAST Team  \\
\small EURATOM/CCFE Fusion Association, Culham Science Centre, \\
\small Abingdon, Oxfordshire OX14 3DB, United Kingdom \\
M Becoulet, F Orain \\
\small Association Euratom/CEA, CEA Cadarache, IRFM, F-13108, St. Paul-lez-Durance, France \\
WA Cooper \\
\small CRPP, Association EURATOM/Conf\'{e}d\'{e}ration Suisse, \\\small EPFL, 1015 Lausanne, Switzerland \\
S Pamela \\
\small IIFS-PIIM. Aix Marseille Universit\'{e} - CNRS, 13397 Marseille Cedex20,France}

\begin{document}
\maketitle

\begin{abstract}

Type-I Edge Localised Modes (ELMs) have been mitigated in MAST through the application of $n=3,4$ and 6 resonant magnetic perturbations (RMPs).
For each toroidal mode number of the non-axisymmetric applied fields, the frequency of the ELMs has been increased significantly, and the peak heat flux on the divertor plates reduced commensurately.
This increase in ELM frequency occurs despite a significant drop in the edge pressure gradient, which would be expected to stabilise the peeling-ballooning modes thought to be responsible for type-I ELMs.
Various mechanisms which could cause a destabilisation of the peeling-ballooning modes are presented, including pedestal widening, plasma rotation braking, three dimensional corrugation of the plasma boundary and the existence of radially extended lobe structures near to the X-point.
This leads to a model aimed at resolving the apparent dichotomy of ELM control, that is to say ELM suppression occurring due to the pedestal pressure reduction below the peeling-ballooning stability boundary, whilst the reduction in pressure can also lead to ELM mitigation, which is ostensibly a destabilisation of peeling-ballooning modes.
In the case of ELM mitigation, the pedestal broadening, 3d corrugation or lobes near the X-point degrade ballooning stability so much that the pedestal recovers rapidly to cross the new stability boundary at lower pressure more frequently, whilst in the case of suppression, the plasma parameters are such that the particle transport reduces the edge pressure below the stability boundary which is only mildly affected by negligible rotation braking, small edge corrugation or short, broad lobe structures.

\end{abstract}
\maketitle

\section{Introduction}

Tokamak plasmas operating in the high-confinement regime exhibit explosive, quasi-periodic instabilities called Edge Localised Modes (ELMs) \cite{Connor}.
Type-I ELMs are understood to be a manifestation of so-called peeling-ballooning instabilities driven by strong pressure gradients and localised current density at the edge of the plasma \cite{Snyder,Wilson}.
Whilst operating tokamak plasmas in a high confinement regime is desirable to maximise the fusion yield, the resultant ELMs can eject large amounts of energy and particles from the confined region, which in turn could result in damage to plasma facing components \cite{Suttrop}.
Indeed, the control of ELMs in future tokamaks, such as ITER, is essential in order to ensure an acceptable lifetime of plasma facing components \cite{Loarte}.
One method to control ELMs -- presently under consideration for ITER -- is the application of resonant magnetic perturbations (RMPs), which perturb the local magnetic field in the pedestal region at the edge of the confined plasma.
RMPs have been applied to completely suppress Type-I ELMs in DIII-D \cite{Evans2004,Evans2008}, ASDEX Upgrade \cite{Suttrop2011} and KSTAR \cite{Jeon} or to mitigate ELMs -- that is to say increase their frequency and reduce their amplitude -- in MAST \cite{Kirk2011,Kirk2012} and JET \cite{Liang}.

The most widely offered explanation for ELM suppression is that the application of RMPs induces a stochastic magnetic field giving rise to enhanced heat and particle transport, which in turn degrades the pressure gradient in the pedestal such that it is below the level required to trigger an ELM \cite{Evans2004,Evans2008,Jeon,Snyder2012}.
Since the pedestal might be expected to continue to widen, and as it does so reduce the critical pressure to trigger a peeling-ballooning mode \cite{Dickinson}, recently it has been postulated that the presence of a magnetic island near the pedestal top acts to prevent the widening of the pedestal \cite{Snyder2012}.
However, the increase in type-I ELM frequency when RMPs are applied, which would seem to be the result of a \emph{destabilisation} of ELMs, cannot be explained through this mechanism.
Indeed, stability analyses of plasmas exhibiting ELM mitigation due to RMPs typically find that the peeling-ballooning stability margin is greatly enhanced \cite{Saarelma}, contrary to the increase in ELM frequency observed.
This work aims to understand this dichotomy in the empirical effect of RMPs -- either a stabilisation of the ELMs by reducing the pressure gradient or a marked destabilisation \emph{despite} a reduction in pressure gradient.
In order to make extrapolations on the effects of RMPs in ITER it is necessary to have a paradigm which can explain how a destabilisation of ELMs occurs when an applied RMP is slightly off resonance, but yet still causes a reduction in the pedestal density gradient which results in ELM suppression when the field is resonant \cite{Evans2008}.
It is worth noting that in high collisionality plasmas, both DIII-D \cite{Evans2008} and ASDEX Upgrade \cite{Suttrop2011} have shown a suppression of type-I ELMs despite little change in the pedestal pressure, though here we concentrate on the effects reported at ITER-relevant low collisionalities.

There are various effects of RMPs manifest in tokamak plasmas:
The most notable one is the reduction in the pedestal density, and therefore, pedestal pressure, which results in a small reduction in the plasma confinement.
However, RMPs can also cause a braking of the plasma rotation \cite{Kirk2011,Garofalo,Sun,Liang2010}, a three-dimensional toroidal displacement -- here referred to as a toroidal \emph{corrugation} -- of the plasma boundary \cite{Chapman2012,Canik} and the introduction of lobe structures near to the magnetic X-point \cite{Kirk2012,Wingen}.

In this paper we consider whether any of these effects could influence the peeling-ballooning edge stability, and so provide insight into the ELM mitigation (ie destabilisation) mechanism.
In section \ref{sec:expt} we briefly present some examples of the experimental phenomenology observed in MAST plasmas when RMPs are used to mitigate ELMs.
Sections \ref{sec:rotation}, \ref{sec:3d} and \ref{sec:lobes} investigate the effects that the plasma rotation braking, the three dimensional distortion of the plasma boundary and the introduction of lobes near the X-point could have on plasma edge stability respectively.
Finally, this leads to the proposal of a model for how RMPs control ELMs in section \ref{sec:model} which unifies the seemingly juxtaposed phenomena of ELM mitigation and ELM suppression.
Conclusions and implications of this work are discussed in section \ref{sec:conclusions}.

\section{Effect of Resonant Magnetic Perturbations in MAST H-modes} \label{sec:expt}

\begin{figure}
\begin{center}
\includegraphics[width=0.5\textwidth]{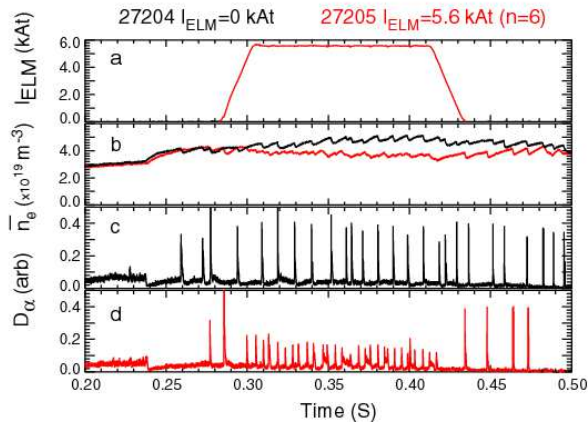}
\end{center}
\caption{The (a) coil in the in-vessel coils, (b) line-averaged electron density, (c) the $D_{\alpha}$ emission in MAST discharge 27205 without RMPs and (d) the $D_{\alpha}$ emission in MAST discharge 27204 with maximum $n=6$ RMP applied. A clear three-fold increase in the ELM frequency, and a 10\% decrease in the plasma density is caused by the RMPs.}
\label{fig:timetraces}
\end{figure}

MAST is equipped with 18 in-vessel ELM control coils capable of applying $n=3,4,6$ non-axisymmetric fields, where $n$ is the toroidal periodicity of the field.
ELM mitigation has been observed with $n=3$ in double null plasmas \cite{Kirk2012b} and $n=3,4,6$ in single null plasmas \cite{Kirk2012,Kirk2012a}, though ELM suppression has not yet been attained.
The single null plasmas, as well as being the most ITER-relevant magnetic configuration, also allow a comparison between $n=3,4,6$ since the plasma is vertically displaced to be closer to the in-vessel coils, maximising the applied field.
A typical example of the ELM mitigation observed with the application of an $n=6$ RMP field is shown in figure \ref{fig:timetraces}.
The divertor $D_{\alpha}$ emission is shown as a function of time for a shot without applied RMPs and one with an $n=6$ field, together with the in-vessel coil currents to demark the time when RMPs are on and the line-averaged density.
It is evident that the application of the non-axisymmetric field results in an increase in ELM frequency by a factor of three, coupled to a reduction in the plasma density.
As the ELM frequency increases, the plasma energy released with each ELM decreases and the peak heat flux on the divertor plates drops commensurately (here by a factor of 1.5 for the plasmas shown in figure \ref{fig:timetraces}).
The ELMs remain characteristic of type-I ELMs in that the frequency increases with injected power and the filament behaviour is identical to that without RMPs \cite{Kirk2012a}.
Previous analysis of ELM filament structures in MAST and ASDEX Upgrade have shown that type-I ELMs have very a different toroidal mode number distribution and range of toroidal propagation speeds compared to small ELMs or Type-III ELMs \cite{KirkAUG}. 
Mitigated ELMs retain exactly the filamentary characteristics of type-I ELMs \cite{Kirk2012a} suggesting their behaviour must still be explained by a stability model for type-I ELMs.
Conversely, analysis of peeling-ballooning stability of mitigated ELMs in DIII-D led to the conclusion that the mitigated ELMs were not type-I ELMs \cite{Hudson}, though this stability analysis is based upon an axisymmetric plasma without lobes near the X-point which we will show in this paper is not an appropriate model when RMPs are applied.
In single-null plasmas the application of RMPs also results in rapid global plasma braking, whilst the plasma rotation is largely unaffected by $n=3$ fields in double null plasmas (which exhibit much faster initial rotation).

\begin{figure}
\begin{center}
\includegraphics[width=0.5\textwidth]{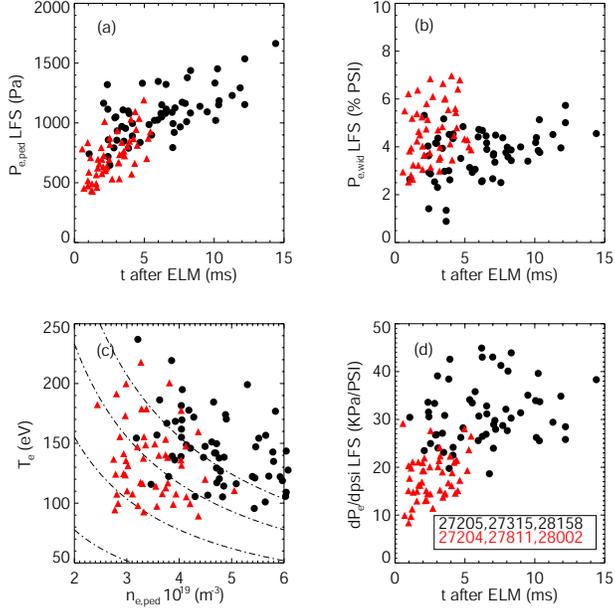}
\end{center}
\caption{The (a) electron pressure at the pedestal top, (b) electron pressure pedestal width in flux space, (c) the electron temperature against the electron density and (d) the electron pressure gradient as a function of time after the previous ELM for a series of MAST shots both without applied RMPs and when an $n=6$ field is applied. The data in the first 10\% of the ELM cycle is ignored. The RMPs cause a reduction in density, leading to a decrease in pressure and pressure gradient, as well as a significant increase in the pedestal width.}
\label{fig:pedestal}
\end{figure}

The pedestal profile evolution when RMPs are applied is diagnosed using MAST's Thomson scattering system which measures the electron temperature and density with a spatial resolution better than 10mm on both the high- and low-field sides of the plasma.
The application of RMPs results in a decrease of the pedestal pressure gradient and an increase of the pedestal width.
Whilst the pedestal evolution does depend upon the RMP configuration applied, this increase in the pedestal width and reduction of the pressure gradient is regularly observed in MAST plasmas when ELM mitigation is observed \cite{Scannell2012}.
Figure \ref{fig:pedestal} shows the electron pressure at the pedestal top, the pedestal width, the electron temperature versus the electron density and the pressure gradient as measured by the Thomson scattering diagnostic for various times since the previous ELM in two identical MAST discharges with and without an applied $n=6$ field.
It is clear that whilst the pedestal pressure evolves in the same way, including starting from the same pedestal height after an ELM, it never reaches the same level when RMPs are applied.
Similarly, the pedestal widens significantly with the application of an $n=6$ field. 
The reduction in pressure and pressure gradient comes predominantly from a reduction in the electron density.

\begin{figure}
\begin{center}
\includegraphics[width=0.5\textwidth]{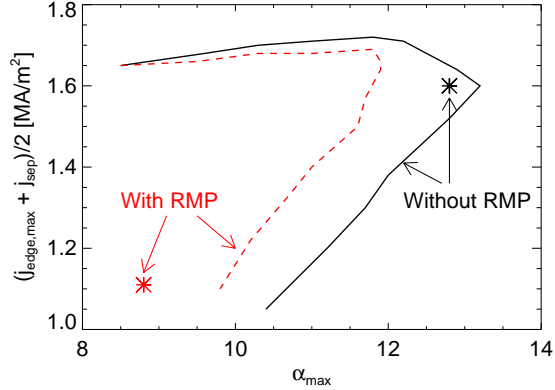}
\end{center}
\caption{The edge stability diagram constructing by varying edge pressure, $\alpha$ and current density, $j$ and reconstructing many different equilibria and testing stability to $n=5,10,15,25,30$ modes. The stability boundary is assessed when the mode growth rate drops below $\gamma/\omega_{A}=0.01$, though the qualitative boundary is unaffected when the marginal point is taken at zero growth rate, or below half of the ion diamagnetic frequency \cite{Saarelma2012}. The star represents the experimental equilibria and the boundary using the pressure profile with and without $n=4$ RMPs has been assessed for a MAST single-null plasma.}
\label{fig:jalpha}
\end{figure}

The broader pedestal does decrease the critical pressure gradient required for peeling-ballooning instability \cite{Dickinson}.
However, the reduced pressure gradient when RMPs are applied is significantly below this new critical threshold, and so the application of RMPs of all toroidal mode numbers enhances stability.
This is demonstrated by stability calculations in figure \ref{fig:jalpha}, where the stability boundary for a MAST single-null discharge with an $n=4$ RMP applied is illustrated.
This stability diagram is constructed by reconstructing the experimental equilibrium and systematically varying the edge pressure gradient and current density around this experimental point.
For each normalised pressure gradient, $\alpha$, and current density, $j$, a new equilibrium is created with the \textsc{Helena} code \cite{Huysmans} and its stability to various finite-$n$ peeling-ballooning modes is tested using the \textsc{Elite} code \cite{Snyder,Wilson2002}. 
This method is described in detail in reference \cite{Saarelma2009}.
Here the plasma shape is taken from \textsc{Efit} \cite{Lao} reconstruction and held fixed, whilst the electron temperature and density profiles are taken from Thomson scattering measurements before and after the application of RMPs, and assumed to be equal to the ion temperature.
Whilst this assumption may be invalid, assuming a flat ion temperature in the pedestal instead has been shown to have little effect on the stability boundary \cite{Dickinson}.
The edge current density is found from a self-consistent bootstrap current iterative calculation constrained by the total plasma current, using the formulae from references \cite{Sauter1999,Sauter2002}.
Having tested finite-$n$ linear stability for $n=5,10,15,20,25,30$ a stability boundary is drawn in this $[j,\alpha]$ space, as illustrated in figure \ref{fig:jalpha}. 
Here, the normalised pressure is defined as
\begin{equation}
\alpha = -\frac{2 \partial V/\partial \psi}{4\pi^{2}} \bigg( \frac{V}{2 \pi^{2} R_{0}} \bigg)^{1/2} \mu_{0} \frac{\partial p}{\partial \psi}
\end{equation}
where $V$ is the volume enclosed by flux surface with poloidal flux, $\psi$, $p$ is the pressure and $R_{0}$ is the major radius of the geometrical plasma centre.
It is clear that when an $n=3$ RMP is applied, the decrease in the observed pressure gradient is expected to make the plasma more stable to finite-$n$ peeling ballooning modes, whilst empirically the type-I ELMs are destabilised and more frequent.

\section{The effect of rotation braking on edge stability} \label{sec:rotation}

\begin{figure*}
\hspace{-0.2cm}
\vspace{0.5cm}
\begin{minipage}{0.33\textwidth}
\begin{center}
\includegraphics[width=\textwidth]{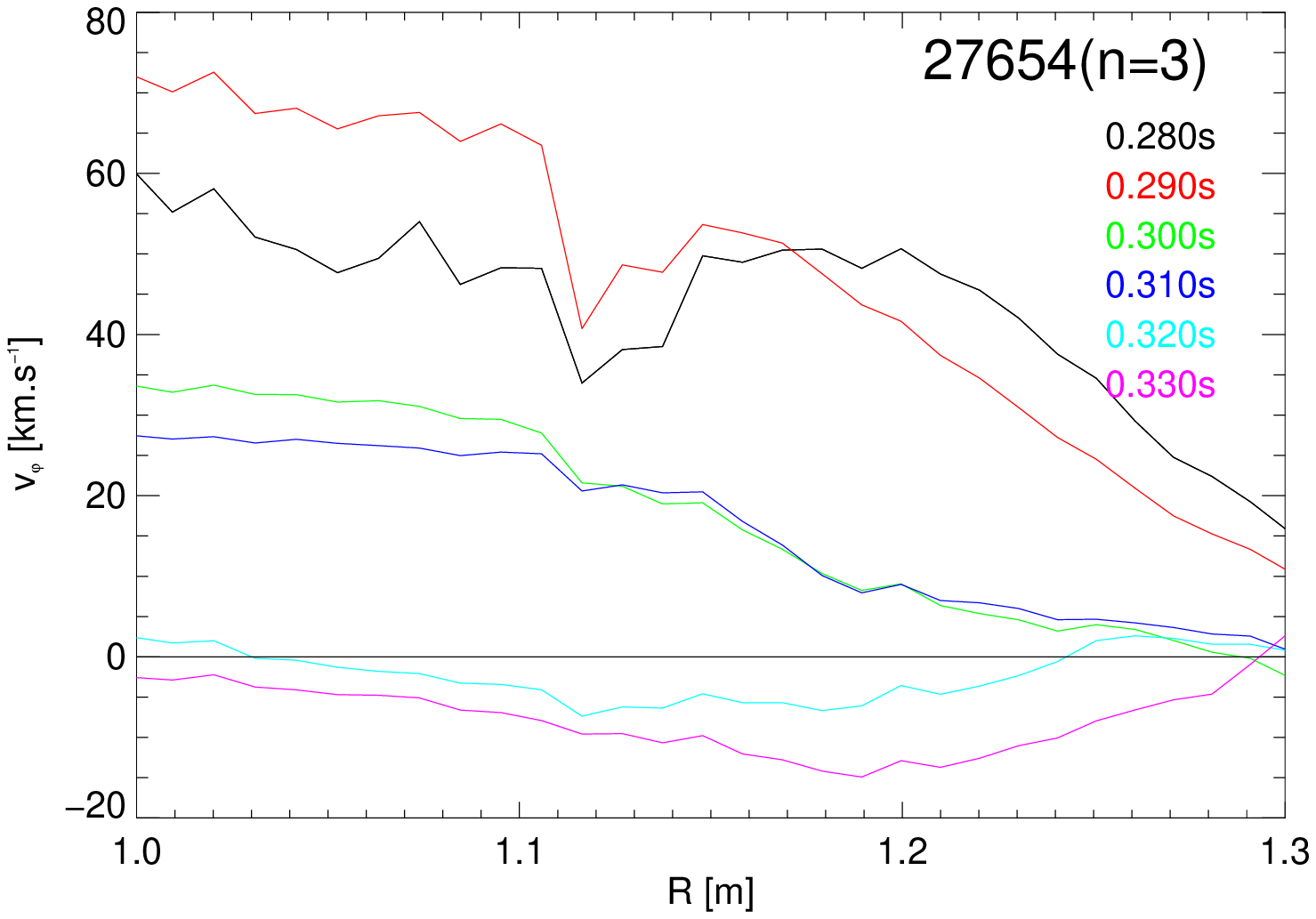}
\end{center}
\end{minipage}
\begin{minipage}{0.33\textwidth} 
\begin{center}
\includegraphics[width=\textwidth]{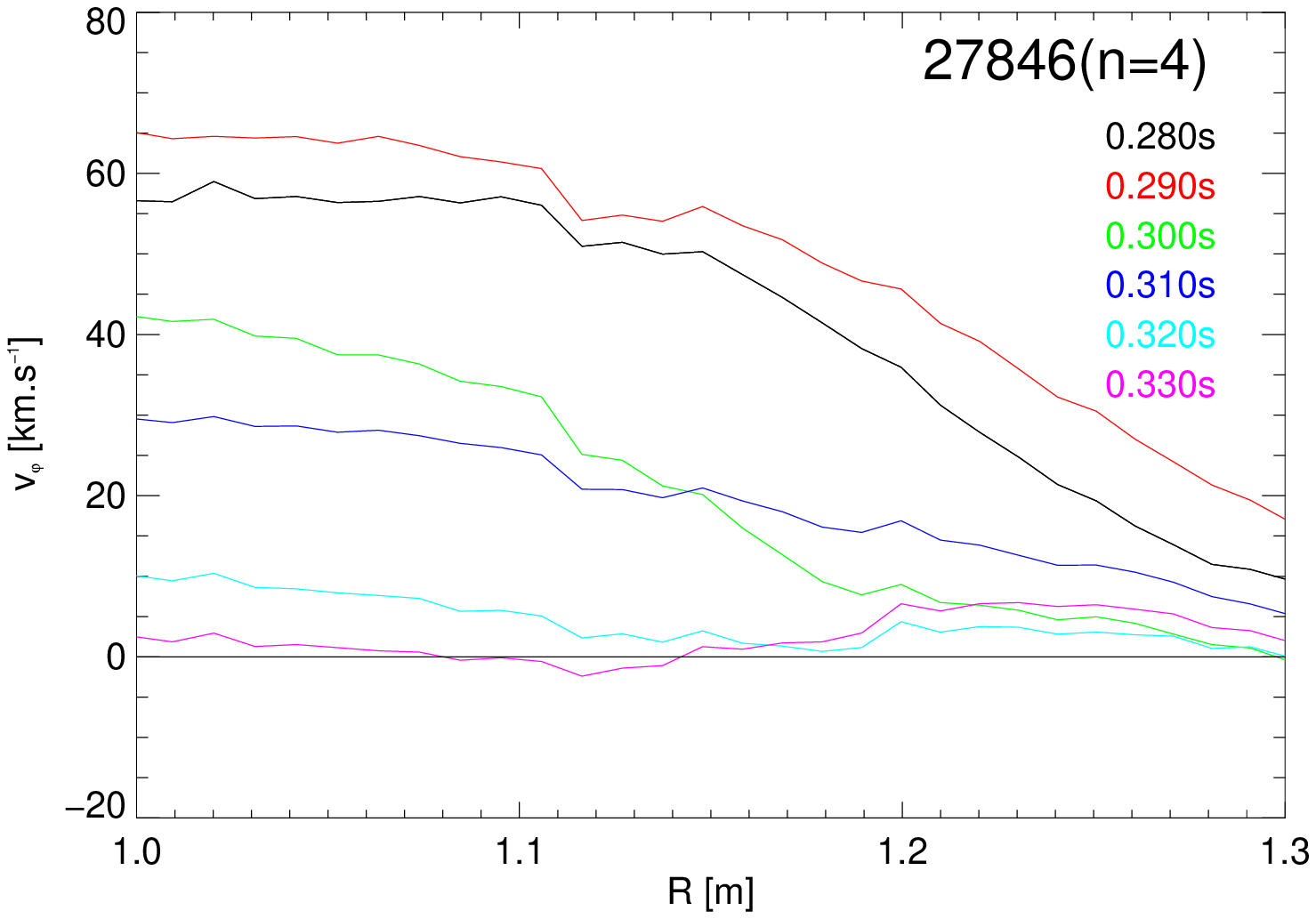}
\end{center}
\end{minipage}
\begin{minipage}{0.33\textwidth}
\begin{center}
\includegraphics[width=\textwidth]{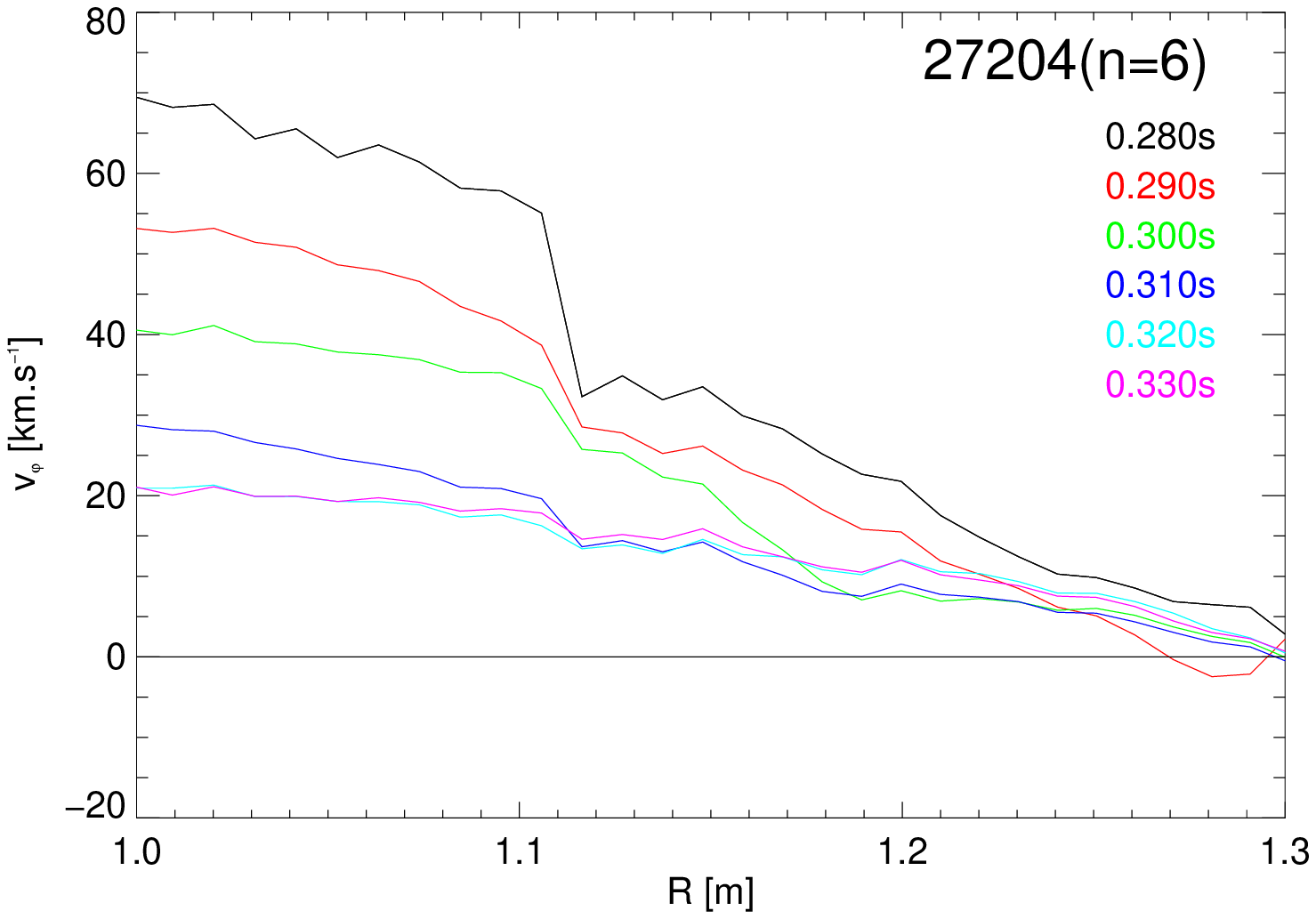}
\end{center}
\end{minipage}
\caption{The radial profile of the toroidal rotation velocity as measured by charge exchange recombination spectroscopy in 10ms time intervals for MAST discharges 27654 ($n=3$ RMP), 27846 ($n=4$ RMP) and 27204 ($n=6$ RMP). In each case the RMP field is turned on at 0.28s and reaches flat-top by 0.3s.}
\label{fig:rotation}
\end{figure*}

When RMPs are applied to single null MAST plasmas, the plasma rotation is observed to decrease, markedly in many cases.
Indeed, for $n=3$ fields at maximum amplitude, the plasma rotation reduces so severely that it causes a back transition to L-mode, and ultimately a plasma disruption.
Figure \ref{fig:rotation} shows the radial profiles of the toroidal rotation velocity measured by charge exchange recombination spectroscopy for different timeslices after $n=3,4,6$ RMPs are applied in typical MAST single-null plasmas.
In all cases there is significant, and global braking of the bulk rotation, with the $n=3$ case locking at zero flow, the $n=4$ RMP causing saturation below 10kms$^{-1}$ and the $n=6$ RMPs resulting in a saturated profile peaking at 20kms$^{-1}$.

\begin{figure*}
\hspace{-0.2cm}
\vspace{0.5cm}
\begin{minipage}{0.33\textwidth}
\begin{center}
\includegraphics[width=\textwidth]{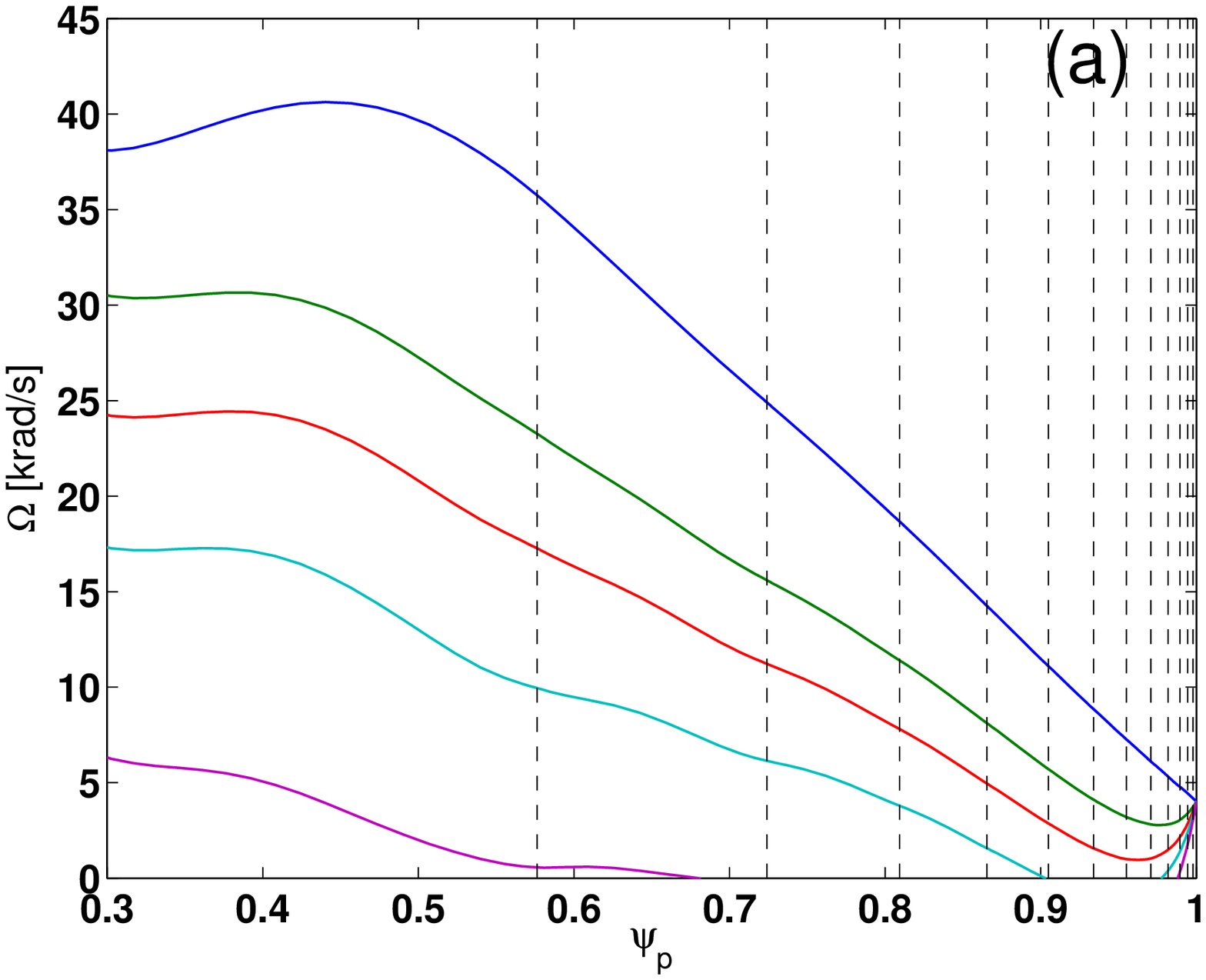}
\end{center}
\end{minipage}
\begin{minipage}{0.33\textwidth} 
\begin{center}
\includegraphics[width=\textwidth]{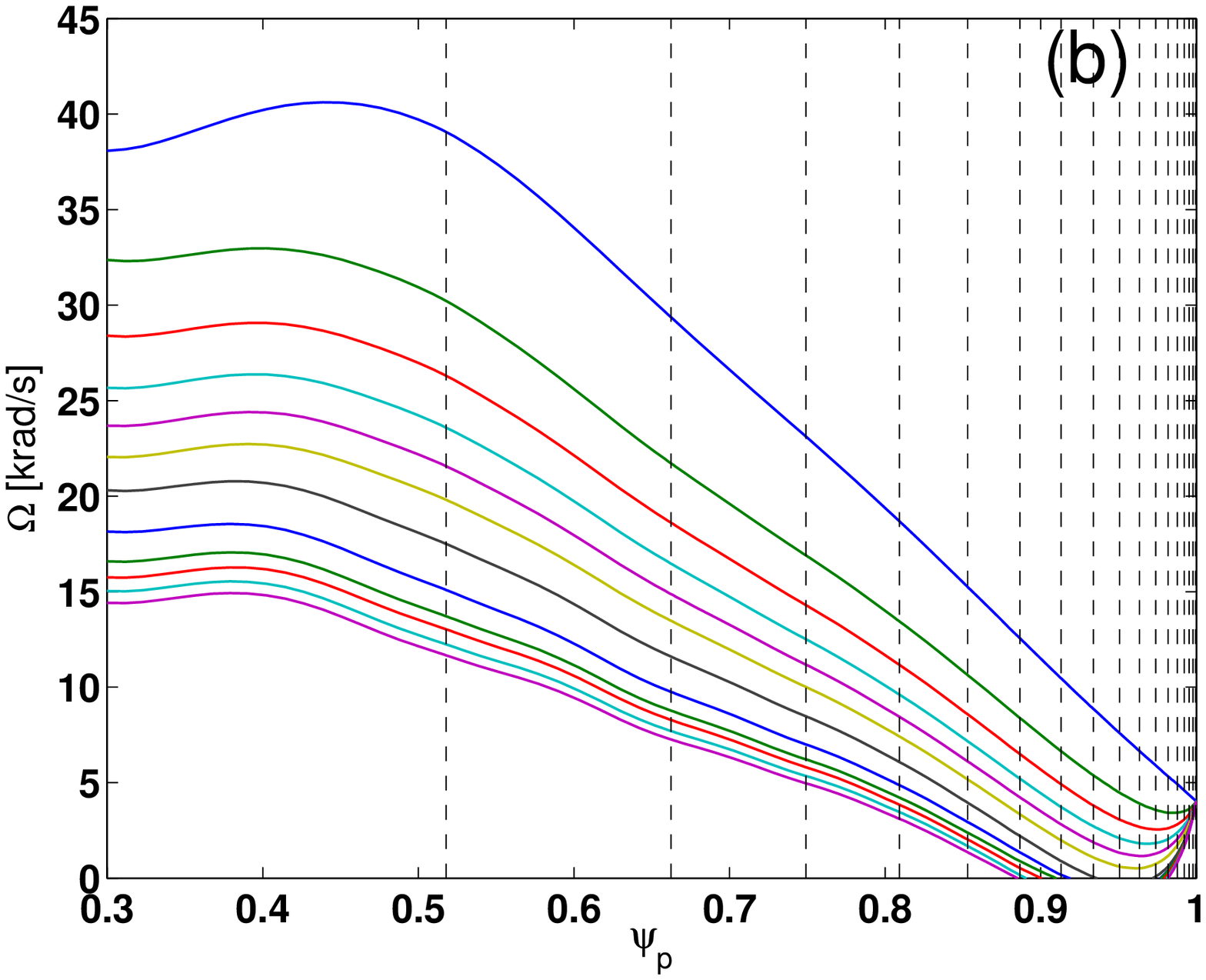}
\end{center}
\end{minipage}
\begin{minipage}{0.33\textwidth}
\begin{center}
\includegraphics[width=\textwidth]{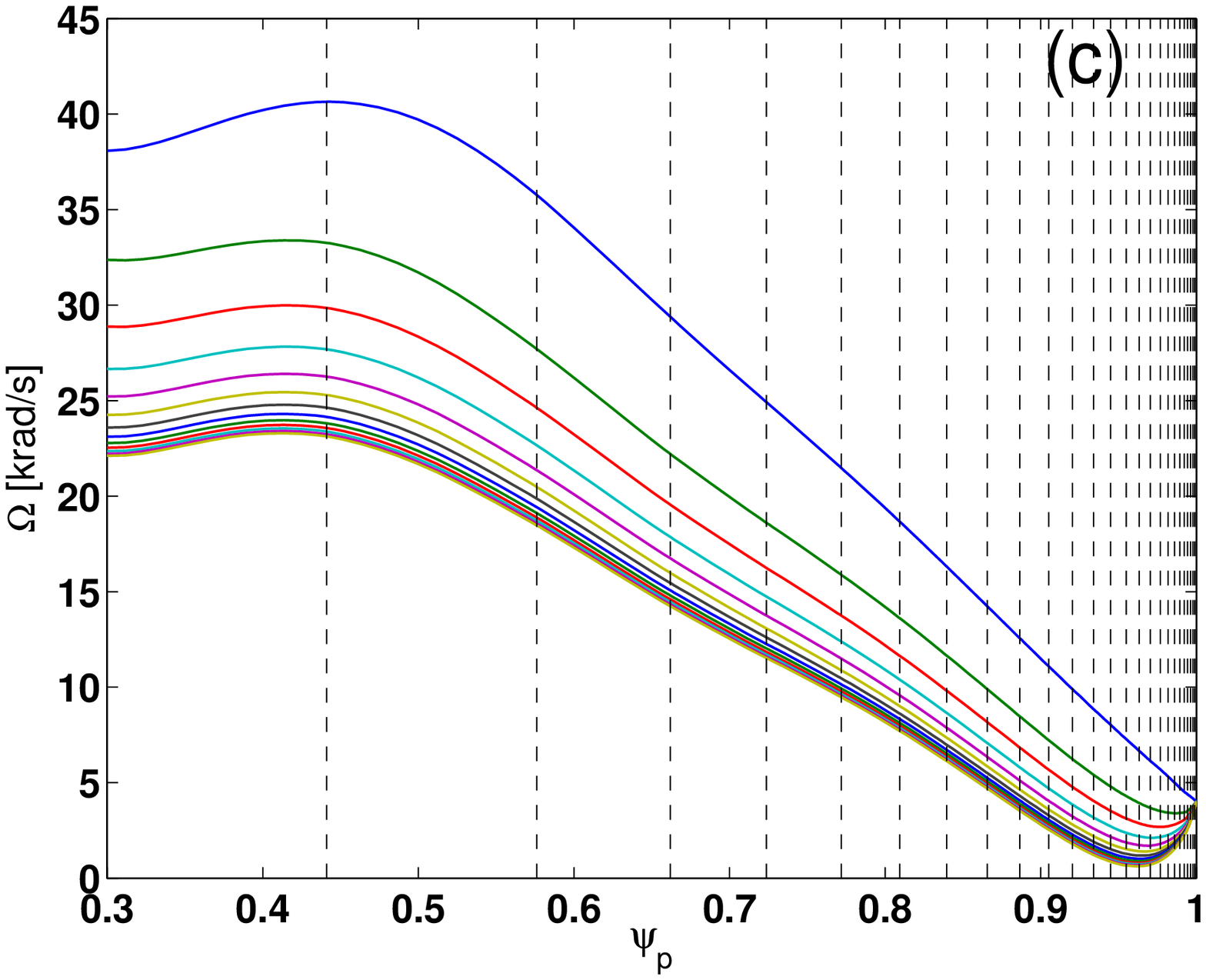}
\end{center}
\end{minipage}
\caption{The radial profile of the toroidal rotation frequency as simulated by MARS-Q in 10ms time intervals for MAST discharges (a) 27654 ($n=3$ RMP), (b) 27846 ($n=4$ RMP) and (c) 27204 ($n=6$ RMP). In each case the edge rotation is fixed throughout. The vertical dashed lines are the rational surfaces.}
\label{fig:marsq}
\end{figure*}

The effect of RMPs on the plasma flow has been simulated using the quasi-linear MARS-Q code \cite{Liu2012}.
MARS-Q employs a full MHD single fluid model for the plasma response, in full toroidal geometry.
The perturbed MHD equations are coupled to a momentum balance equation for the change in the toroidal flow of the plasma and solved using a semi-implicit adaptive time-stepping scheme non-linearly in time.
The model includes both the $\textbf{j} \times \textbf{B}$ torque \cite{Fitzpatrick} and the NTV torque \cite{Shaing}, with both resonant and non-resonant contributions included.
To simulate the MAST plasmas shown in figure \ref{fig:rotation}, a resistive plasma model in assumed with magnetic Lundquist number $S_{0}=3.5 \times 10^{7}$ at the magnetic axis, with a profile given by $S \sim T_{e}^{3/2}$, so $S_{edge} \approx 10^{6}$. 
Figure \ref{fig:marsq} shows the radial profiles of the toroidal rotation at timeslices approximately 10ms apart.
Only the radial region outside $\psi=0.3$ is illustrated for comparison to figure \ref{fig:rotation} since the line of sight of the charge exchange diagnostic does not reach the axis in these vertically displaced lower single-null plasmas.
In 40ms the $n=3$ RMP causes a complete global braking (noting that the edge boundary condition to keep the edge rotation fixed preclude it going to zero), whilst the $n=4$ and $n=6$ saturate at finite values, qualitatively similar to the experiment, and in good agreement with the temporal evolution too.
In all cases, when there is initially fast plasma rotation, the $\textbf{j} \times \textbf{B}$ torque is dominant.
However, as the rotation is damped by the $n=3$ field and approaches full damping, the NTV torque becomes dominant and the $\textbf{j} \times \textbf{B}$ torque vanishes.
In the $n=4$ case the two torques become comparable at saturation whilst the $\textbf{j} \times \textbf{B}$ torque remains dominant throughout the rotation evolution when the $n=6$ RMP is applied. 
The penetrated field is significantly larger at the end of the nonlinear evolution since the screening rotation is removed, though it is worth noting that it remains markedly smaller than the vacuum field treatment.
Finally, as the rotation decreases and the applied field penetrates further, the perturbed MHD displacement increases and becomes more peaked near the X-point.

\begin{figure}
\begin{center}
\includegraphics[width=0.5\textwidth]{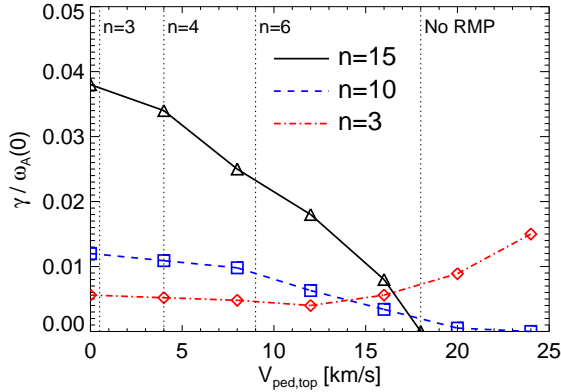}
\end{center}
\caption{The growth rate of $n=3$, $n=10$ and $n=15$ peeling-ballooning modes as a function of rotation velocity at the pedestal top. The rotation profile is fixed and takes a modified tanh profile shape across the pedestal region. Also shown are the saturated pedestal-top rotation speeds when $n=3,4,6$ RMPs are applied compared to the initial rotation velocity.}
\label{fig:eliteflow}
\end{figure}

The reduction in plasma rotation both in MAST and in MARS-Q simulations is global and drastic.
Previously it has been shown that the edge flow shear can stabilise ballooning modes \cite{Saarelma2007,Xi} and drive peeling modes \cite{Aiba,Xi,Snyder2007}.
Therefore it is of interest to simulate whether the changes in the rotation profile resultant from the application of RMPs affect ELM stability.
In order to simulate this, the \textsc{Elite} linear MHD code \cite{Snyder,Wilson2002} has been used, taking the rotation profile as a tanh profile in the pedestal and the amplitude of the pedestal top rotation from the experimental measurements.
The region with sheared flow profile has a width of 5\% of the poloidal flux and is centred at the maximum pressure gradient ($\psi=0.985$).
\textsc{Elite} takes a static equilibrium, and so can only be considered as accurate at low flow speeds \cite{Chapman2010b,Wahlberg}.
Figure \ref{fig:eliteflow} shows the growth rate of $n=3,10,15$ peeling-ballooning modes as a function of the pedestal top rotation speed, with the initial rotation speed, and the final saturated speed under the influence of $n=3,4,6$ RMPs marked.
Here the pedestal pressure has been scaled so that the equilibrium is marginally stable at the initial rotation velocity.
The decrease in the rotation observed when RMPs are applied is predicted to destabilise the high-$n$ ballooning modes, yet slightly stabilise the low-$n$ peeling modes.
When the rotation is slowed to the level observed under application of $n=3$ fields, the growth rate of (the most unstable) mid-$n$ peeling-ballooning mode is massively increased, whereas for higher-$n$ RMPs, the destabilisation is weaker.

However, there is little evidence of pedestal braking measured experimentally, and indeed, the toroidal propagation of the filaments after each ELM is approximately the same with and without RMPs applied \cite{KirkIAEA}, suggesting that the pedestal rotation is unchanged.
Furthermore, there are examples of single-null plasmas with strong core rotation braking but little affect on the ELM frequency, as well as double null plasmas which exhibit no change in rotation profile, but an ELM frequency which approaches an order of magnitude greater than without RMPs.
This suggests that the braking of the rotation, whilst marked and global in some single null MAST plasmas, is sub-dominant in determining the ELM behaviour.

\section{The effect of three-dimensional corrugation of the plasma boundary on edge stability} \label{sec:3d}

\begin{figure}
\begin{center}
\includegraphics[width=0.4\textwidth]{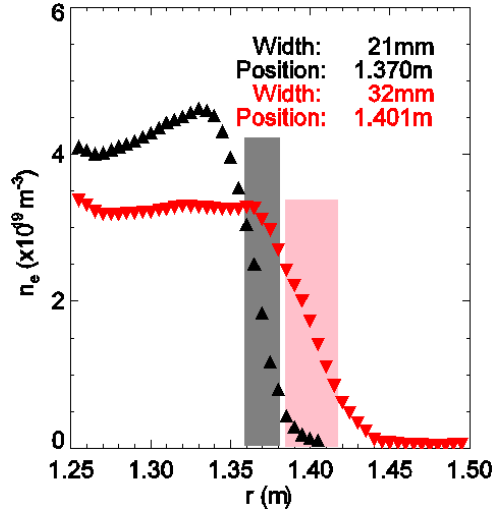}
\end{center}
\caption{The electron density radial profile in the pedestal region as measured by the Thomson scattering diagnostic for discharges with and without an $n=6$ RMP applied. When the RMP is applied (red line), the pedestal width clearly increases and the position of the outboard midplane moves out by approximately 5cm. The inboard position is unaffected.}
\label{fig:TS}
\end{figure}

MAST's 18 in-vessel coils means that various phases of each toroidal mode number of magnetic perturbation can be applied.
Furthermore, MAST is equipped with a number of diagnostics which can measure the outboard position of the plasma with sub-centimeter radial resolution, making it possible to accurately measure the corrugation resultant from the application of RMPs.
In reference \cite{Chapman2012}, the boundary corrugation is considered for two phases of an $n=3$ field with a 60$^{\circ}$ phase shift.
In this case, various diagnostics measured a corrugation of approximately 2cm from the case without a non-axisymmetric field, giving rise to $\sim$4cm shift between the two opposite phases of the $n=3$ field.
This large displacement of the plasma boundary measured in various toroidal locations is equivalent to more than 5\% of the minor radius for an applied field amplitude necessary to mitigate the ELMs.
An example of the displacement of the plasma boundary as measured by the Thomson scattering diagnostic is shown in figure \ref{fig:TS}, where a large shift of the plasma edge for the case with an $n=6$ RMP can be seen.

\begin{figure}
\begin{center}
\includegraphics[width=0.6\textwidth]{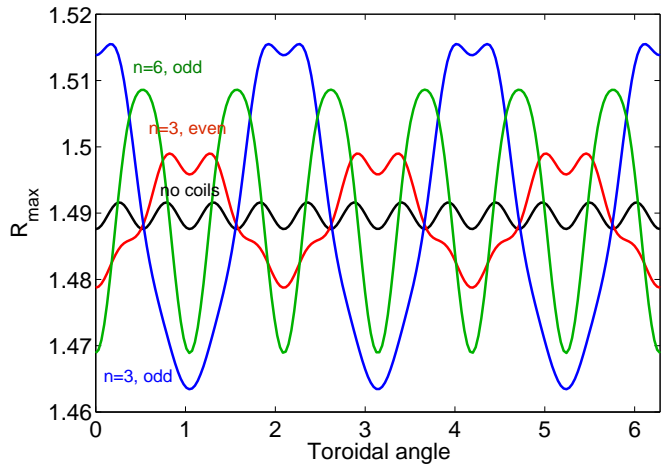}
\end{center}
\caption{The midplane boundary as a function of toroidal angle as modelled by the \textsc{Vmec} free-boundary 3d equilibrium code for different applied $n=3,6$ fields in MAST. The boundary shows a clear $n=3,6$ periodic corrugation in addition to the negligible $n=12$ toroidal field ripple. The displacement is maximised when the parity of the applied field is resonant with the equilibrium $q$-profile.}
\label{fig:animec}
\end{figure}

Such a large boundary corrugation clearly requires a non-axisymmetric treatment.
The measured corrugation of the plasma edge reported in \cite{Chapman2012} can be compared with the numerical reconstruction of a non-axisymmetric plasma equilibrium using the \textsc{Vmec} code \cite{Hirshman}.
Instead of solving the Grad-Shafranov equation to find an equilibrium state, \textsc{Vmec} uses a variational method to find a minimum in the total energy of the system.
The major radius of the plasma edge at the midplane for various amplitudes of applied fields as predicted by the free-boundary version of \textsc{Vmec} is shown in figure \ref{fig:animec}.
In the absence of an RMP there is a imperceptibly small $n=12$ corrugation due to the toroidal field ripple.
As the current in the in-vessel coils is increased to the maximum (5.6kAt), the $n=3$ corrugation linearly increases.
At full field, the corrugation is approximately 5cm between the extrema, in good agreement with that measured experimentally, despite the absence of islands or field screening in the simulation.
The boundary displacement is dependent upon the alignment between the applied field and the equilibrium field.
For an even parity case (when the current in upper and lower coils has the same sign), the corrugation is significantly less than for odd parity $n=3$ fields.
The corrugation incurred by $n=6$ RMPs, whilst greater than 5\% of the minor radius, is less than for an optimal $n=3$ RMP.

\begin{figure}
\begin{center}
\includegraphics[width=0.6\textwidth]{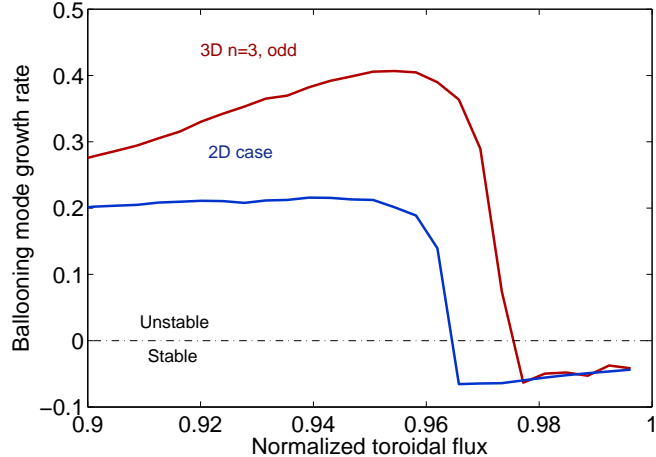}
\end{center}
\caption{The infinite-$n$ ballooning mode stability parameter as a function of toroidal flux, focusing on the pedestal region, for an axisymmetric case and for the most unstable toroidal position when an odd-parity $n=3$ RMP is applied. The application of the RMPs leads to a 3d corrugation of the plasma boundary which in turn leads to increased ballooning mode drive in certain toroidal locations.}
\label{fig:infinite_n}
\end{figure}

The influence of this three dimensional corrugation on infinite-$n$ ballooning stability has also been examined using the \textsc{Cobra} code \cite{Sanchez}.
Not only does the 3d field change the local curvature with an $n=3$ periodicity, it also leads to a piling-up and rarefaction of pressure surfaces along the corrugation.
Previously an $n=1$ kink was found to strongly drive ballooning modes in certain toroidal positions, whilst stabilising the $n=\infty$ ballooning modes in others, as the corrugation caused a compression and rarefaction of pressure surfaces in different toroidal locations \cite{Chapman2010}.
Figure \ref{fig:infinite_n} shows the $n=\infty$ ballooning mode stability parameter as a function of radius, where positive growth rate indicates instability.
The axisymmetric treatment finds that the pedestal region is infinite-$n$ unstable, as expected given the strong pressure gradient in this region.
The growth rate of the $n=\infty$ ballooning modes at the most unstable toroidal position is a factor of two larger than the axisymmetric case, showing that the nonaxisymmetry strongly destabilises the plasma edge in certain toroidal positions.
Recent two-dimensional electron cyclotron emission imaging of ELMs on KSTAR has shown that there is a strong toroidal asymmetry of ELM filaments \cite{Yun}.
Each ELM filament is seen to occur as discrete bursting fingers at different toroidal locations.
This implies that a mechanism that changes the local ballooning stability at any given toroidal location would affect ELM behaviour.

\section{The effect of lobe structures near the X-point on edge stability} \label{sec:lobes}

\begin{figure}
\begin{center}
\includegraphics[width=\textwidth]{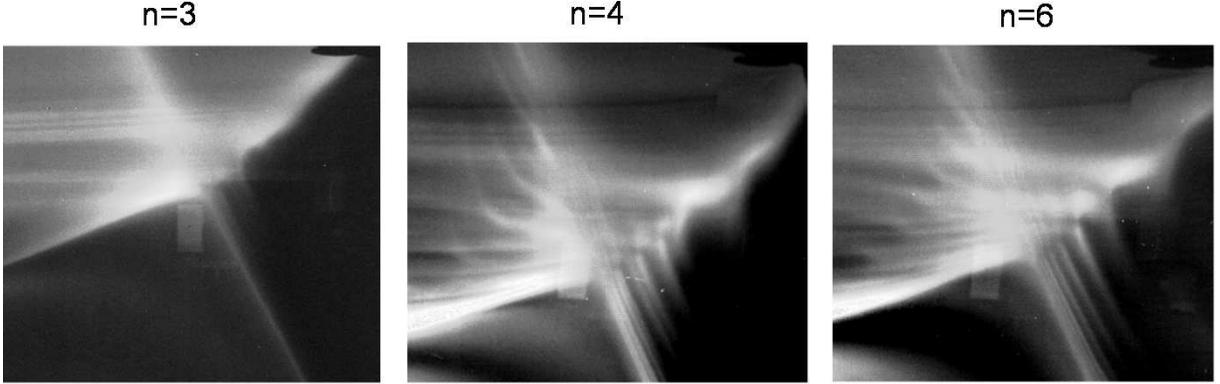}
\end{center}
\caption{High-speed visible camera images obtained with a HeII filter near the X-point during an inter-ELM period of H-modes in MAST when a)$n=3$, b)$n=4$ and c)$n=6$ RMPs are applied.}
\label{fig:lobes}
\end{figure}

The lower X-point region of MAST single null plasmas can be imaged using a toroidally viewing camera with spatial resolution of 1.8mm in the tangency plane, filtered with either a HeII (468nm) or CIII (465nm) filter and using a 300$\mu$s integration time.
Figure \ref{fig:lobes} shows images obtained with the HeII filter during an inter-ELM period when $n=3,4,6$ RMPs are applied to a MAST H-mode.
A deformation of the separatrix is observed and clear lobe structures can be seen near to the X-point \cite{Kirk2012}.
The location, radial extent and poloidal separation of the lobes is different for each toroidal mode number of the applied field \cite{Harrison}.
The lobe structures are only observed above a critical applied field threshold.
This threshold is consistent with the critical non-axisymmetric field required to affect the plasma, that is to say, for enhanced particle transport (so-called density pump-out) to be observed in L-mode or for an increase in ELM frequency to be observed in H-mode.

It is known that introducing an arbitrarily small, radially and poloidally localised bump (used to approximate the effect of an X-point) on the low field side of tokamak plasmas degrades ballooning stability, whilst vastly enhancing peeling stability \cite{Bishop,Saarelma2011,Huysmans2005,Webster2009}. 
If the lobe is on the low-field side of the torus, then the plasma is more susceptible to ballooning instabilities  since the field-lines exist for much of their length in the region of unfavourable curvature ($\langle \kappa \cdot \nabla P \rangle <0$), whilst approaching the X-point when the poloidal field tends to zero.
From reference \cite{Saarelma2011} we know that the ballooning stability is progressively degraded the further onto the low-field side the lobe is positioned.
To examine the edge stability when lobe structures are present in the separatrix after application of resonant magnetic perturbations, an axisymmetric stability analysis using the \textsc{Elite} code has been performed.
The analysis presented here necessarily has nested flux surfaces with no magnetic islands or stochastisation of the edge field and assumes the lobes are axisymmetric. 
There is no doubt that this is a gross simplification of the experimental situation, but the concepts may help to understand how ELM control by resonant magnetic perturbation occurs.

\begin{figure}
\begin{center}
\includegraphics[width=0.6\textwidth]{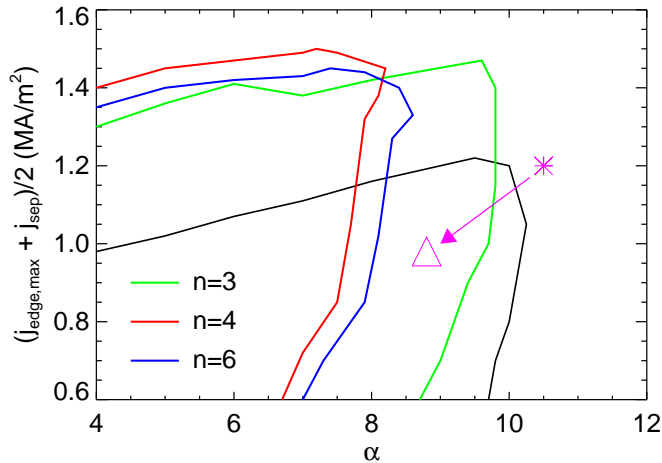}
\end{center}
\caption{Finite-$n$ peeling-ballooning stability boundaries for a MAST single null plasma with axisymmetric lobes present as observed experimentally under application of $n=3,4,6$ RMPs. The star represents the experimental equilibrium prior to an ELM before applying the RMPs, with the triangle the operational parameters after RMPs are applied.}
\label{fig:lobe_stability}
\end{figure}

Figure \ref{fig:lobe_stability} shows the peeling-ballooning boundaries for plasma shapes that are achievable in a 2d equilibrium code and most akin to those seen in the visible imaging (figure \ref{fig:lobes}) when RMPs are applied.
Here, the $j-\alpha$ diagram is constructed using the method outlined in section \ref{sec:expt}.
It is evident that the presence of lobes degrades the ballooning stability boundary significantly, whilst at the same time, stabilises low-$n$ peeling modes.
This degradation in ballooning stability originates from the perturbed field lines dwelling in the region of unfavourable curvature due to the presence of lobe structures rather than the change in the plasma boundary shape.
The axisymmetric treatment employed here forces the poloidal field in the lobes to become very small, which in turn causes the destabilisation of the ballooning modes \cite{Chapman2012b}.
The strongest effect from $n=4$ shown in figure \ref{fig:lobe_stability} is due to having a combination of a lobe further on the low-field side than either $n=3$ or $n=6$ and importantly having at least two narrow lobes.
The $n=3$ perturbation has least effect since the lobes are more poloidally extended, so do not increase the magnetic shear nor degrade the curvature as much.
Of course, the position and poloidal extent of these three-dimensional lobe structures would be different in a different toroidal plane, and so have a subtly different effect on the stability boundary.
An accurate treatment of the effect of the RMP-induced lobes on edge stability requires a full three-dimensional analysis, and this is the subject of future work.

\section{A model for the effect of RMPs on ELM Stability} \label{sec:model}

\begin{figure}
\begin{center}
\includegraphics[width=0.6\textwidth]{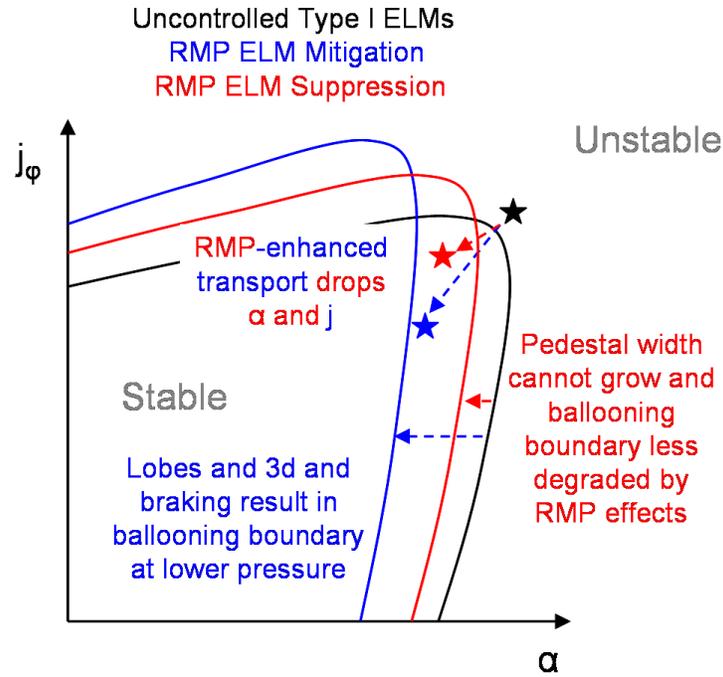}
\end{center}
\caption{A model for how RMPs affect ELM behaviour illustrated in peeling-ballooning stability space, viz. current density against normalised pedestal pressure gradient. In a typical type I ELMing plasma, an ELM is triggered when the pressure and current profiles (black star) reach the corner of the stability boundary (black line). When RMPs are applied, the enhanced particle transport leads to a reduction in the pressure, and commensurate reduction in the pedestal bootstrap current. In the case of RMP mitigation, the combined effect of RMP-induced plasma braking, 3d corrugation of the plasma boundary and lobes near the X-point is to significantly degrade ballooning stability, indicated by the ballooning boundary moving to lower normalised pressure gradients (blue line). The pedestal recovers to this lower stability boundary more rapidly after the previous ELM, and so the ELM frequency increases. In the case of ELM suppression, the ballooning boundary is not as degraded (red line), and the RMP-induced particle transport means that the operational point now sits in the stable region, hence an absence of ELMs.}
\label{fig:j-alpha-sketch}
\end{figure}

It is evident that as well as dropping the pedestal pressure gradient, the RMPs have various effects which influence the edge stability: (1) a change in the edge rotation shear \cite{Liu2012,Rozhansky,Tokar}, which is known to affect peeling-ballooning modes \cite{Saarelma2007}; (2) the 3d corrugation of the plasma boundary \cite{Chapman2012}, since it has been shown that ballooning modes can be more unstable in local toroidal positions when the plasma is non-axisymmetric \cite{Bird,Chapman2010}; (3) steady-state lobe structures existing in the unfavourable curvature region near the X-point \cite{Chapman2012b}; or (4) a broadening of the pressure pedestal.
When RMPs are applied, the edge pressure gradient is usually reduced \cite{Liang,Kirk2011,Evans2008} (though in ASDEX Upgrade it is barely affected \cite{Suttrop2011}), meaning that the operational point moves towards a more peeling-ballooning stable region in the conventional $j-\alpha$ space used to parametrize ELM stability (see, for instance, refs \cite{Snyder,Wilson,Saarelma}).
Typically, after an ELM, the pedestal evolution results in a broadening of the pedestal \cite{Dickinson} until the peeling-ballooning boundary is crossed since a wider pedestal means a lower critical pressure for instability.
This happens since lower-$n$ modes are usually most unstable, but can only grow when the pedestal is sufficiently extended that the mode eigenfunction can fit within the pedestal region where the driving pressure gradient is localised.
However, if the plasma parameters (edge flows, $q-$profile resonance with respect to the applied field, density gradients and so on) are such that the pedestal saturates at a width which gives a marginal stability boundary above the actual pressure gradient determined by RMP-induced particle transport, then ELM suppression occurs.
Conversely, if the destabilisation afforded by any, or all, of the change in edge rotation shear, 3d corrugation or lobe structure deformation is so strong that the marginal stability boundary is sufficiently degraded to a point below the operational pressure gradient, then the ELMs are destabilised.
Furthermore, since the pedestal relaxes to a similar pressure immediately after an ELM with and without RMPs, as illustrated in figure \ref{fig:pedestal}, the pedestal recovers to the degraded stability boundary more rapidly, and therefore the ELM frequency increases, resulting in ELM mitigation.
There is no reason per se that suppression requires a stronger degradation of the ballooning stability boundary; indeed, perhaps the optimal situation is one whereby there is only a small degradation in ballooning stability, but sufficient pedestal transport that the pedestal will never reach this degraded stability boundary.
This paradigm is sketched in figure \ref{fig:j-alpha-sketch}.
It should be noted that this model is based purely on results from linear stability analysis, whilst the inter- and intra-ELM evolution is very much a nonlinear process.

\section{Conclusions} \label{sec:conclusions}

Various mechanisms which can cause a degradation in edge stability when RMPs are applied have been considered.
Ballooning modes are found to be more unstable when the edge flow shear is reduced, when the plasma configuration exhibits a three dimensional boundary corrugation and when lobe structures exist near the X-point, all of which are a result of the application of RMPs.
These destabilisation mechanisms begin to explain how ELM mitigation occurs when RMPs are applied, despite a reduction in the pedestal pressure and pressure gradient which would be expected to result in a more stable plasma edge.
Consequently the different ELM control effects -- either ELM mitigation manifest as more frequent, smaller ELMs, or ELM suppression -- can both be explained by an RMP-induced reduction of the pedestal pressure.
It is hypothesised that the former occurs in the case where the RMP effects also result in a significant degradation in the ballooning stability boundary, whilst ELM suppression may occur when the stability boundary is only slightly affected and the RMP-induced transport lower the pedestal parameters below this marginal stability point (noting that a mechanism which precludes pedestal broadening is required to avoid a lower-$n$ peeling-ballooning mode eventually being destabilised).
Each of these effects play a role in peeling-ballooning stability, so optimising ELM control requires a subtle balance between them all.
For instance, figures \ref{fig:rotation} and \ref{fig:marsq} show that in MAST the $n=3$ RMPs cause stronger braking than $n=6$ but figure \ref{fig:lobes} shows that the radial extent of the lobes is bigger for an $n=6$ field than $n=3$ RMP.
Furthermore, these effects are connected in a nonlinear way; as an example, it is experimentally observed that as the plasma rotation is reduced by the RMPs, the lobe length increases, presumably as the rotation screening of the applied field is diminished \cite{Harrison}, which would exacerbate the destabilisation of the peeling-ballooning mode.

Using this paradigm, it may become possible to assess the implications for ITER and discriminate between the efficacy of various RMP configurations, although more complete numerical modelling is required to make any quantitative assessment of peeling-ballooning stability (ie 3d stability of finite-n peeling-ballooning modes, treating the lobes as helical rather than axisymmetric structures, etc).
If ELM suppression is required to allow operation at 15MA compatible with acceptable divertor plate lifetime, then a configuration which induces least rotation braking (noting that the flow in ITER is predicted to be small anyway), least boundary corrugation and shortest lobe structures is desirable.
Conversely, if high-frequency ELM pacing is more desirable, for instance to help with flushing high-Z impurities from the plasma \cite{Romanelli,Loarte}, then it is best to use a coil configuration which maximises these effects.
To minimise the rotation braking, the RMP configuration should be non-resonant to reduce the $j \times B$ torque and the poloidal spectrum should be set-up in such a way as to reduce the NTV torque.
To minimise the edge corrugation, the applied field should be non-resonant and the plasma profiles ought to be far from marginal stability for resonant low-$n$ peeling modes, which will otherwise amplify the applied field and enhance the distortion. 
This can be achieved by varying the edge safety factor, and even the shear through localised current drive, to minimise the peeling mode drive.
Lastly, the lobe length can be reduced either by using a non-resonant field, or by using lower-$n$ fields which typically produce less radially extended and more poloidally extended lobes.
Of course, the RMP-field itself must still be sufficiently large to incur enhanced particle transport to prevent the pedestal reaching the stability boundary.

\noindent \textbf{Acknowledgements}

\noindent This work was partly funded by the RCUK Energy Programme under grant EP/I501045 and the European Communities under the contract of Association between EURATOM and CCFE. The views and opinions expressed herein do not necessarily reflect those of the European Commission. 


\pagebreak


\end{document}